\documentclass[prl,twocolumn,amsmath,amssymb,superscriptaddress,longbibliography]{revtex4-2}
\usepackage{xcolor}
\usepackage{graphicx}
\usepackage{hyperref}
\usepackage{oplotsymbl}

\begin{document}

\title{Extreme outbreaks in non-Markovian epidemics on complex networks}

\author{Ami Taitelbaum}
\affiliation{Racah Institute of Physics, The Hebrew University, 9190401 Jerusalem, Israel}
\author{Michael Assaf}
\affiliation{Racah Institute of Physics, The Hebrew University, 9190401 Jerusalem, Israel}

\date{\today}

\begin{abstract}

Extreme epidemic risk is controlled by the right tail of the outbreak-size distribution,
but this distribution is generally unknown for non-Markovian spreading on networks.
Here we determine this distribution by mapping non-Markovian SIR dynamics to an effective Markovian description.
We show that arbitrary infection and recovery time statistics can be incorporated through a single edge transmissibility,
yielding an effective Markovian process that reproduces the full outbreak-size statistics.
For weakly heterogeneous networks,
the reduction yields a universal well-mixed semiclassical theory governed by the bond-percolation reproductive number.
Outbreak statistics across diverse waiting-time distributions and topologies collapse onto one predictive curve.
For highly heterogeneous and empirical networks,
the corresponding effective Markovian dynamics on the network captures the complete distribution.
Our results provide a direct route from measured waiting-time distributions to quantitative predictions of network-level extreme-outbreak risk.
\end{abstract}

\maketitle

Compartmental epidemic models, notably the susceptible--infected--recovered (SIR) model~\cite{Kermack1927}, underpin mathematical epidemiology~\cite{Anderson1991,Keeling2008}.
In finite populations, stochastic fluctuations generate a broad distribution of epidemic outcomes that is not captured by deterministic mean-field theory.
Because epidemic risk depends on both fade-out probabilities~\cite{Britton2010} and the upper tail of outbreak sizes~\cite{House2013,CirilloTaleb2020}, risk assessment requires the full outbreak-size distribution, not only a typical final outbreak size.
For example, even with a moderate mean outbreak, there remains a small but consequential risk of overwhelming hospital capacity~\cite{parnass2023estimating}.

For finite well-mixed populations, the final outbreak-size distribution can be computed recursively~\cite{clancy2014sir,Miller2019}, and its large-deviation form has been obtained~\cite{Hindes2022,hindes2023outbreak} using a semiclassical (WKB) approach~\cite{dykman1994large,assaf2010extinction,AssafMeerson2017}.
In this framework, the probability of an outbreak of fractional size $x_r^*$ scales as
$P(x_r^*) \sim e^{-N\mathcal{S}(x_r^*)}$, where $\mathcal{S}$ is a large-deviation action controlled by the epidemic parameters and initial condition.
On networks, the topology provides an additional source of complexity. Percolation approaches determine thresholds and deterministic final sizes~\cite{Grassberger1983,Newman2002,Miller2011}, while stochastic network results are largely restricted to Gaussian fluctuations~\cite{Ball2021} or Markovian SIS extinction rates~\cite{hindes2019degree}.
Varying the network topology is expected to dramatically change not only the deterministic final size, but the entire outbreak-size distribution including its rare tails.
Currently, a large-deviation theory for the full non-Gaussian outbreak-size distribution on heterogeneous networks remains unavailable.

Non-Markovian epidemics arising from nonexponential waiting-time (WT) distributions render the dynamics more intricate and create substantial modeling and analytical challenges~\cite{Kiss2015,DiDomenico2024,ascione2026interplay}. Notably, such nonexponential distributions are commonly observed for empirical generation intervals and infectious periods~\cite{Lloyd2001a,Wearing2005,Ferretti2020}.
Pairwise closures, generalized mean-field theory, and message passing characterize thresholds and time courses for non-Markovian spreading~\cite{KarrerNewman2010,VanMieghem2013,Cator2013,Kiss2015,RostViziKiss2018,VanMieghem2019}, while first-passage approaches compute transmission statistics on networks~\cite{GiuggioliSarvaharman2022}.
For well-mixed non-Markovian epidemics, transmissibility-based final-size relations can also be derived~\cite{MaEarn2006}. Related equivalence approaches treat non-Markovian SIS steady states on networks~\cite{Starnini2017,Feng2019}.
Recently, continuous-time random-walk and memory-kernel methods~\cite{aquino2017chemical,vilk2024non} were applied to extend large-deviation theory to non-Markovian well-mixed epidemics~\cite{ShmunikAssaf2026}.
However, a theory combining network topology and non-exponential WTs at the level of the full outbreak-size distribution, including extreme events, remains elusive.

Here we reveal a universal final outbreak-size distribution across diverse WT distributions and network topologies, and show that its normalized shape can be predicted by employing an effective well-mixed, Markovian theory~\cite{Hindes2022}.
For weakly-heterogeneous networks, the bond-percolation reproductive number determines the position on this universal curve, enabling quantitative prediction of the full distribution.
For strong heterogeneity the universality breaks down; yet, the per-edge transmissibility~\cite{Newman2002} still maps the non-Markovian dynamics onto effective Markovian dynamics \textit{on the same network structure}, capturing the outbreak-size distribution.
We also apply this formalism to empirical social networks with assortative mixing, such as the Hamsterster network~\cite{Kunegis2013}, enabling quantitative predictions for the risk of unusually large outbreaks in real-life scenarios.

Consider the SIR model with $N$ individuals, each in one of three states: susceptible ($S$), infected ($I$), or recovered ($R$), such that $S+I+R=N$. Transitions occur through two processes: infection and recovery. In a well-mixed setting (fully-connected network) the probability per unit time that the number of susceptibles decreases by one and the number of infecteds increases by one is $\beta S I$, $\beta$ being the per-connection infection rate, whereas the probability per unit time that the number of infecteds decreases by one is $\gamma I$, $\gamma$ being the recovery rate.
Denoting by $x_w = W/N$ the population fraction in state $W\in\{S,I,R\}$, for fully-connected networks the mean-field dynamics satisfy
\begin{equation}
    \dot{x}_s = -\beta N\,x_s x_i,\quad \dot{x}_i = \beta N\,x_s x_i - \gamma x_i,\quad \dot{x}_r = \gamma x_i,
\end{equation}
where this result is valid under Markovian dynamics, see below. At this point, it is useful to define the basic reproduction number, $R_0 = \lambda N$, where $\lambda \equiv \beta/\gamma$ is the per-edge infection-to-recovery ratio.
For $R_0 > 1$ the final recovered fraction $x_r^* \equiv x_r(t\to\infty)$ has a deterministic mean $\overline{x_r^*}$ satisfying $1 - \overline{x_r^*} = e^{-R_0\,\overline{x_r^*}}$~\cite{Kermack1927,Anderson1991}.

We now generalize this well-mixed Markovian picture to non-Markovian dynamics with arbitrary WT distributions on finite complex networks.
Consider $N$ nodes on an undirected network with degree distribution $p_k$---the fraction of nodes having $k$ neighbors---and mean degree $\bar{k} = \sum_k k\,p_k$. Initially, we consider uncorrelated networks and later we will extend the analysis and include assortative mixing~\cite{Newman2002assort}.
In real epidemics, infection and recovery durations are not memoryless: generation intervals and infectious periods exhibit variability captured by measured distributions~\cite{Ferretti2020,Wearing2005}.
We model this by assigning each active infection or recovery channel an independent clock drawn from a general WT distribution.
Each infected node can transmit the disease to each susceptible neighbor, with infection WT distribution $f_i(t)$ and recovery WT distribution $f_r(t)$, where the corresponding means are set to be $1/\beta$ and $1/\gamma$, respectively.
The Markovian epidemic threshold is $\lambda_c^{-1} = \sigma_k^2/\bar{k} + \bar{k} - 2$~\cite{PastorSatorras2015}, where $\sigma_k$ is the standard deviation of the degree distribution $p_k$.
On networks, $R_0 \equiv \lambda/\lambda_c$ generalizes the well-mixed definition, recovering $\lambda N$ for a large complete network in which $\lambda_c \to 1/N$.

To proceed, we define the per-edge transmissibility~\cite{Newman2002}
\begin{equation}\label{eq:T}
T = \int_0^{\infty} f_i(t)\, \Phi_r(t)\, dt
\end{equation}
as the probability that infection along one S--I edge occurs before recovery, where $\Phi_r(t) = \int_t^{\infty} f_r(t')\,dt'$ is the recovery survival function.
Within the SIR model, infection along each edge occurs at most once; the final state is therefore a bond-percolation configuration with occupation probability~$T$~\cite{Newman2002,Grassberger1983}.
For a single initial infected node, the final infected cluster is the connected component of that node in the subgraph of occupied edges.
On a locally tree-like network, a node of degree $k$ escapes infection only if none of its $k$ edges transmit infection to it, giving an escape probability $\theta^k$, where $\theta$ is the probability that infection does not reach the node through a given edge.
Defining the probability generating function of the degree distribution, $G_0(x) = \sum_k p_k x^k$~\cite{NewmanStrogatzWatts2001},
the mean outbreak fraction can be calculated through the bond-percolation self-consistency equations~\cite{Newman2002,Miller2011}
\begin{equation}\label{eq:perc_gen}
\overline{x_r^*} = 1 - G_0(\theta), \quad \theta = 1 - T + T\,G_0'(\theta)/\bar{k}.
\end{equation}
Notably, for Erd\H{o}s--R\'enyi (ER) networks~\cite{ErdosRenyi1959}, Eq.~\eqref{eq:perc_gen} simplifies to $\overline{x_r^*} = 1 - e^{-\bar{k} T\, \overline{x_r^*}}$~\cite{NewmanStrogatzWatts2001}, recovering the well-mixed final-size equation with $\bar{k} T$ playing the role of $R_0$.

Observed generation intervals and infectious periods are commonly modeled by gamma distributions~\cite{Ferretti2020,Wearing2005}.
As a prototypical example, we therefore take both $f_i$ and $f_r$ to be gamma-distributed:
\begin{equation}\label{gammadist}
   f(t) = \frac{(\alpha r)^{\alpha}}{\Gamma(\alpha)}\,t^{\alpha-1}\,e^{-\alpha r t}. 
\end{equation}
Here $\Gamma(z)=\int_0^\infty t^{z-1}e^{-t}\,dt$ is the gamma function, and the parameters $\alpha$ and $r$ satisfy $(\alpha,r)=(\alpha_{\rm inf},\beta)$ for infection and $(\alpha_{\rm rec},\gamma)$ for recovery, with mean $r^{-1}$ and standard deviation $r^{-1}/\sqrt{\alpha}$ (we henceforth set $\gamma=1$ and measure time in units of the mean recovery period).
The shape parameter $\alpha$ controls the width of the WT distribution: $\alpha=1$ recovers the exponential (Markovian) limit, $\alpha<1$ gives a broader-than-exponential distribution, while $\alpha>1$ gives rise to a narrower one.
Plugging this distribution [Eq.~(\ref{gammadist})] into Eq.~\eqref{eq:T} yields
\begin{equation}\label{eq:T_general}
T = I_z(\alpha_{\rm inf},\,\alpha_{\rm rec}),\quad
z = \frac{\alpha_{\rm inf}\beta}{\alpha_{\rm inf}\beta + \alpha_{\rm rec}\gamma},
\end{equation}
where $I_z(a,b) = B_z(a,b)/B_1(a,b)$ is the regularized incomplete beta function and $B_z(a,b)=\int_0^z t^{a-1}(1-t)^{b-1}\,dt$ is the incomplete beta function.
Equation~\eqref{eq:T_general} simplifies in special cases: for exponential recovery ($\alpha_{\rm rec}=1$) it gives $T = [\alpha_{\rm inf}\beta / (\alpha_{\rm inf}\beta + \gamma)]^{\alpha_{\rm inf}}$, and for exponential infection ($\alpha_{\rm inf}=1$) it gives $T = 1 - [\alpha_{\rm rec}\gamma / (\alpha_{\rm rec}\gamma + \beta)]^{\alpha_{\rm rec}}$.
Both cases reduce to $\beta/(\beta+\gamma)$ in the fully Markovian limit ($\alpha_{\rm inf}=\alpha_{\rm rec}=1$)~\cite{Miller2011}.

\begin{figure}[t]
\includegraphics[width=1.03\columnwidth]{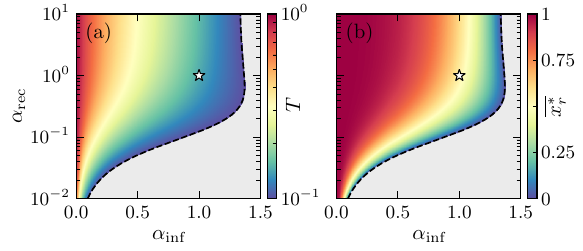}
\vspace{-7mm}\caption{\label{fig:heatmap}
(a)~Per-edge transmissibility $T$ (log-scale colorbar)
versus shape parameters $\alpha_{\rm inf}$ and $\alpha_{\rm rec}$, for an ER network with $\bar{k} = 10$ and $R_0 = 1.5$, with gamma-distributed infection and recovery. 
(b)~Deterministic outbreak fraction $\overline{x_r^*}$ for the same network.
Dashed line: epidemic threshold; grey region: sub-threshold regime; star: Markovian point ($\alpha_{\rm inf} \!=\! \alpha_{\rm rec} \!=\! 1$).}
\end{figure}

Figure~\ref{fig:heatmap} shows the transmissibility $T$ and the deterministic outbreak fraction $\overline{x_r^*}$ as functions of the shape parameters $\alpha_{\rm inf}$ and $\alpha_{\rm rec}$.
The dependence is markedly asymmetric: because $T$ involves the infection density $f_i$ but the recovery \emph{survival} function $\Phi_r$, broader-than-exponential infection ($\alpha_{\rm inf} < 1$) increases $T$ while broader-than-exponential recovery ($\alpha_{\rm rec} < 1$) decreases it.
Moreover, the sensitivity differs strongly: $\alpha_{\rm inf}$ controls the outbreak fraction across its full range, driving near-complete outbreaks for $\alpha_{\rm inf} < 1$ and pushing the system below threshold for large $\alpha_{\rm inf}$. On the contrary, $\alpha_{\rm rec}$ has a sizable effect only for $\alpha_{\rm rec} \ll 1$.
We have numerically checked (not shown) that similar trends are observed for Weibull and log-normal WT distributions.

\begin{figure}[t]
\includegraphics[width=.86\columnwidth]{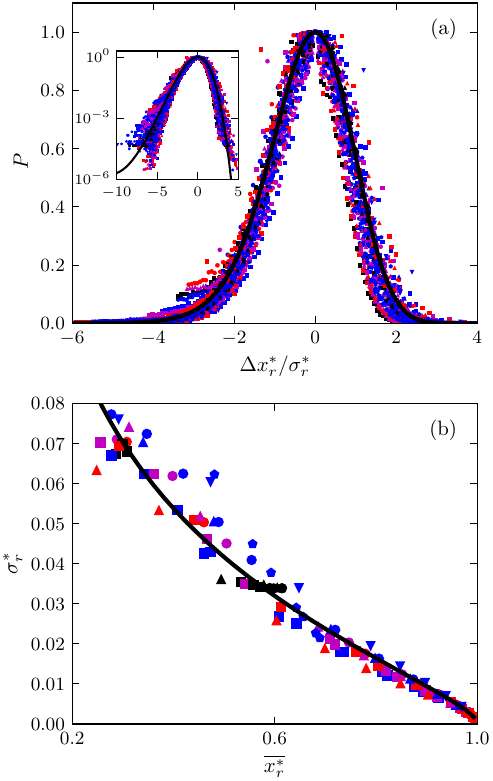}
\vspace{-4mm}\caption{\label{fig:collapse}Universal collapse of outbreak statistics for non-Markovian epidemics on networks  with $N=3000$, $\sigma_k/\bar{k} < 0.5$, and $R_0 = 1.2$ and $1.5$.
The markers encode the network type and WT distribution: circles ($\circ$), squares ($\square$) and triangles ($\triangle$) are regular, ER,  and gamma networks with gamma-distributed WTs for infection and exponential recovery. Inverted triangles ($\triangledown$) and pentagons ($\pentago$) are regular networks with $\bar{k}\!=\!10$, respectively with log-normal WTs for infection (see Appendix~\ref{app:distributions}), and gamma WTs for  recovery.
Color encodes $\bar{k}$ or $\sigma_k$ (blue: smallest, magenta: middle, red: largest), and black markers are Markovian simulations ($\alpha\!=\!1$).
(a)~Rescaled outbreak-size distributions.
Each distribution is centered at its empirical mode and scaled by its conditional standard deviation~\cite{note_peak}.
Solid curve: WKB theory from Eq.~\eqref{eq:action} with $R_0 = 1.5$.
Inset: same data on semi-logarithmic scale.
(b)~Conditional standard deviation $\sigma_r^*$ versus mean $\overline{x_r^*}$.
Solid curve: WKB prediction from Eq.~\eqref{eq:sigma}, parametric in $R_0$.}
\end{figure}

We now go beyond the deterministic outbreak-size and compute the full outbreak-size distribution, including rare extreme events.
Figure~\ref{fig:collapse} shows simulation results for multiple network topologies, mean degrees, and non-Markovian shape parameters; see Appendix~\ref{app:distributions} for details.
For weakly heterogeneous networks (degree coefficient of variation $\lesssim 0.5$), outbreak statistics are remarkably universal: rescaled distributions collapse onto a single curve, as shown in Fig.~\ref{fig:collapse}(a), and the conditional mean and standard deviation follow a single parametric relation, as shown in Fig.~\ref{fig:collapse}(b).
This universality is captured by the well-mixed WKB theory~\cite{dykman1994large,assaf2010extinction,AssafMeerson2017}, which predicts the outbreak-size distribution $P(x_r^*) \sim e^{-N \mathcal{S}(x_s^*)}$~\cite{Hindes2022} with the action
\begin{multline}\label{eq:action}
\mathcal{S}\left(x_s^*\right) = \ln x_s^* + (1-x_s^*)\\
  \times\!\left[m(1+R_0 x_s^*) - 1 + \ln\frac{m(R_0+1)-1}{x_s^*\, m^2 R_0}\right]\!.
\end{multline}
Here, $x_s^* = 1-x_r^*$ and $m$ parameterizes the outbreak trajectory via $m(R_0 x_s^* + 1) - 1 = [m(R_0+1)-1]\,e^{-m R_0(1-x_s^*)}$.
Physically, $m = e^{p_i}$ is the exponentiated momentum conjugate to the infected fraction in the WKB Hamiltonian: $m=1$ (i.e., $p_i=0$) selects the deterministic mean-field trajectory ($\mathcal{S}=0$), while $m \neq 1$ describes rare fluctuations of larger ($m>1$) or smaller ($m<1$) outbreaks.
The Gaussian width near the deterministic peak is~\cite{Hindes2022}
\begin{equation}\label{eq:sigma}
\sigma_r^* = \frac{1}{\sqrt{N\,\mathcal{S}''(\overline{x_s^*})}},
\quad
\mathcal{S}''(\overline{x_s^*}) = \frac{(R_0 \overline{x_s^*} - 1)^2}
           {(1-\overline{x_s^*})\,\overline{x_s^*}\,(R_0^2 \overline{x_s^*} + 1)}.
\end{equation}
The WKB action, Eq.~\eqref{eq:action}, reproduces the rescaled distribution shape, including its non-Gaussian tails, as shown in Fig.~\ref{fig:collapse}(a). Apart from finite-size deviations in the left tail, the right, large-outbreak tail, is universal across network structures and WT shapes. Consistently, the conditional mean and standard deviation obey the universal relation shown in Fig.~\ref{fig:collapse}(b), which is well captured by Eq.~\eqref{eq:sigma}. Together, these results give rise to an effective description of non-Markovian dynamics on complex heterogeneous networks using well-mixed Markovian theory.

\begin{figure}[t]
\includegraphics[width=\columnwidth]{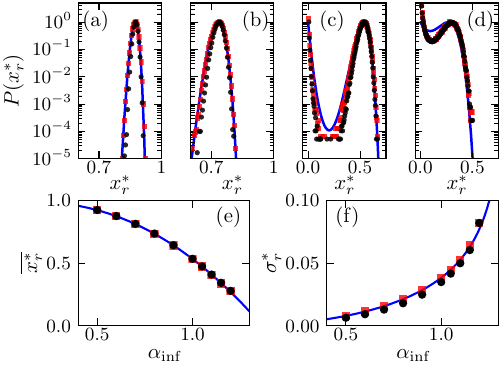}
\vspace{-7mm}\caption{\label{fig:distributions}ER network with $N=3000$, $\bar{k} = 10$, $R_0 = 1.5$, gamma WTs for infection and exponential recovery.
(a)--(d)~Outbreak-size distributions for $\alpha_{\rm inf} = 0.6,\,0.8,\,1.0,\,1.2$, each normalized by its extensive-outbreak peak.
Black circles: non-Markovian network simulations.
Blue line: WKB theory for $R_0=R_0^{\rm eff}$ taken from Eq.~\eqref{eq:R0_eff}.
Red squares: well-mixed Markovian simulation with $N=3000$ for the same $R_0^{\rm eff}$.
(e)~Conditional mean $\overline{x_r^*}$ and (f)~conditional standard deviation $\sigma_r^*$ versus $\alpha_{\rm inf}$.}
\end{figure}

To map the network results onto this well-mixed curve, we invert the deterministic final-size relation $1 - \overline{x_r^*} = e^{-R_0\,\overline{x_r^*}}$ using the network prediction from Eq.~\eqref{eq:perc_gen}:
\begin{equation}\label{eq:R0_eff}
R_0^{\rm eff} = -\ln(1-\overline{x_r^*})/\overline{x_r^*}.
\end{equation}
Substituting $R_0 \to R_0^{\rm eff}$ in Eqs.~\eqref{eq:action}--\eqref{eq:sigma} allows predicting the outbreak-size distribution, as shown in Fig.~\ref{fig:distributions}(a)--(d).
Broader-than-exponential infection times ($\alpha_{\rm inf} < 1$) place more weight at short intervals, raising the chance of transmission before recovery and thus increasing $T$.
Narrower-than-exponential times ($\alpha_{\rm inf} > 1$) have the opposite effect and suppress the outbreak.
As $\alpha_{\rm inf}$ decreases, the conditional mean grows while the standard deviation shrinks, see Fig.~\ref{fig:distributions}(e)-(f), precisely following the universal trend of Fig.~\ref{fig:collapse}(b).
Complementary results for non-exponential recovery are given in Appendix~\ref{app:recovery}.

\begin{figure}[t]
\includegraphics[width=.89\columnwidth]{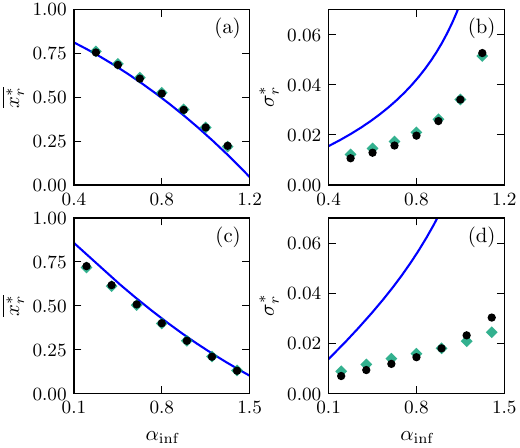}
\vspace{-4mm}\caption{\label{fig:heterogeneous}Conditional mean (left) and standard deviation (right) of the final outbreak-size distribution versus infection shape parameter $\alpha_{\rm inf}$ on highly heterogeneous networks.
Panels (a,b): gamma network with $N=3000$, $\bar{k}=10$, $\sigma_k = 8$, and $R_0 = 1.5$. Panels (c,d): Hamsterster network with $N=2426$, $R_0 = 3$.
Black circles: non-Markovian network simulations with gamma WTs for infection and exponential recovery.
Green diamonds: Markovian (exponential) simulations on the same network at $R_0^{\rm net}$ from Eq.~\eqref{eq:R0_net}.
Blue: WKB with $R_0^{\rm eff}$ from Eq.~\eqref{eq:R0_eff}. Here we have used an extended bond-percolation mapping due to the degree-degree correlations in the Hamsterster network, see Appendix~\ref{app:assortative}.}
\end{figure}

For networks with larger degree heterogeneity, the well-mixed mapping breaks down.
An alternative way to move forward retains the network structure: inverting the Markovian relation, $T = \lambda/(1+\lambda)$~\cite{PastorSatorras2015}, gives~\footnote{When $\bar{k}\gg 1$ and $\sigma_k/\bar{k}\ll 1$, $R_0^{\rm eff}$ reduces to $R_0^{\rm net}$. For networks with $\sigma_k = {\cal O}(\bar{k})$, as occurs in many realistic networks, $R_0^{\rm net}$ cannot replace $R_0^{\rm eff}$ in the mapping to an effective well-mixed model.}
\begin{equation}\label{eq:R0_net}
R_0^{\rm net} = \frac{T}{1-T}\,\lambda_c^{-1}.
\end{equation}
On such networks, effective Markovian simulations at $R_0=R_0^{\rm net}$ track the non-Markovian outbreak statistics, while the well-mixed WKB with $R_0=R_0^{\rm eff}$ fails, as shown in Fig.~\ref{fig:heterogeneous}.
The universal collapse breaks down systematically as degree heterogeneity increases, because achieving the same $\overline{x_r^*}$ on a broader-degree network demands smaller $\alpha_{\rm inf}$, concentrating infection events and narrowing the outbreak-size distribution, see Appendix~\ref{app:heterogeneity} and Fig.~\ref{fig:std_vs_cov}.
For the Hamsterster network~\cite{Kunegis2013,Feng2019}, an empirical social network with broad degree distribution and strong degree-degree correlations, we extend the bond-percolation mapping, see Appendix~\ref{app:assortative}.
Using this generalized percolation relation, the effective Markovian mapping captures the Hamsterster outbreak statistics, as shown in Fig.~\ref{fig:heterogeneous}(c), (d), and  the full distributions, see Appendix~\ref{app:assortative}, Fig.~\ref{fig:hamsterster_dist}.

The per-edge transmissibility $T$ reduces non-Markovian WT statistics to a single effective parameter.
Combined with the network's degree structure, it provides two complementary routes to the final outbreak-size distribution.
For weakly heterogeneous networks, $R_0^{\rm eff}$ maps the problem onto a universal well-mixed WKB theory.
For highly heterogeneous networks, $R_0^{\rm net}$ parameterizes effective Markovian dynamics on the network itself.
Notably, for Markovian recovery and non-Markovian infection, the effective rate $T/(1-T)$ agrees with the SIS effective rate derived in~\cite{Cator2013}.
However, that treatment predicts a shape-independent rate for non-Markovian recovery, which does not hold for SIR.
Shape-dependent SIR results for Markovian infection and non-Markovian recovery on regular networks were obtained in~\cite{Kiss2015,RostViziKiss2018,ViziKissMillerRost2019}.
Our framework unifies these approaches; it combines non-Markovian infection and recovery with arbitrary network topologies, to give the full outbreak-size distribution including extreme  tails.

To demonstrate the power of our method we quantify the risk of extreme outbreaks on two prototypical networks. Conditioning on the occurrence of an extensive epidemic, we compare three ranges relative to the mean final outbreak size $\overline{x_r^*}$: outbreaks lying up to 20\% below the mean, up to 20\% above the mean, and more than 20\% above the mean.
On an ER network, choosing $N=3000$, $\bar{k}=10$, $R_0=1.5$, and $\alpha_{\rm inf}=1.2$ for which $\overline{x_r^*}\approx 0.28$, and $\sigma_r^*\approx0.07$, the corresponding probabilities are approximately 25\%, 32\%, and 21\%.
On the Hamsterster network, choosing $R_0=3$ and $\alpha_{\rm inf}=1.4$ for which $\overline{x_r^*}\approx 0.13$ and $\sigma_r^*\approx0.03$, the corresponding probabilities are approximately 28\%, 34\%, and 19\%.
In both examples, the right tail carries substantial probability mass of about 20\%, indicating a significant risk of unusually large outbreaks.
By connecting measured contact patterns and inter-event times~\cite{Mossong2008,Ferretti2020} to extreme-outbreak probabilities, our framework thus bridges epidemiological data and network-level risk assessment.
Calibrating the mapping to pathogen-specific generation-interval measurements could yield actionable outbreak-risk bounds for emerging infections in real-life scenarios.

We thank M. Shmunik and A. Leibenzon for useful discussions.

\bibliography{references}


\twocolumngrid

\appendix

\setcounter{section}{0}
\newcommand{\appsection}[1]{%
  \refstepcounter{section}%
  \section*{Appendix \Alph{section}: #1}%
}
\setcounter{equation}{0}
\renewcommand{\theequation}{S\arabic{equation}}
\setcounter{figure}{0}
\renewcommand{\thefigure}{S\arabic{figure}}

\appsection{Simulation details}
\label{app:distributions}

This appendix describes the simulation methods, and the different networks and WT distributions used throughout the manuscript.

We simulate the stochastic epidemic dynamics using a Monte Carlo first-reaction method.
Upon activation, each transmission or recovery channel draws a single firing time from its WT distribution.
The earliest event fires, while non-fired channels retain their original times until invalidated.
This is exact because active channels are statistically independent.
For exponential WTs this reduces to the standard stochastic simulation algorithm~\cite{Gillespie1976,Gillespie1977}, while for general WTs it gives an exact simulation of non-Markovian dynamics, verified against the modified next-reaction method of Ref.~\cite{Bogunaetal2014}.
Well-mixed Markovian simulations are performed using the modified next-reaction method~\cite{Anderson2007}.
Each realization is seeded by a single initially infected node chosen uniformly at random, where all others are susceptible.
For each parameter set, $10^6$ independent realizations are performed. For all the synthetic networks,  20 network configurations are created and we run $5\times10^4$ realizations per configuration, while  the Hamsterster network is used as is, see below.
Outbreak statistics are conditioned on extensive outbreaks, defined as realizations above the minimum separating the two modes of the bimodal outbreak-size distribution.

Throughout the manuscript we employ four network families.
(i)~\emph{Regular}: all nodes have degree $k$ exactly ($p_k=\delta_{k,\bar{k}}$, $\sigma_k=0$).
(ii)~\emph{Erd\H{o}s--R\'enyi} (ER): each pair of nodes is connected independently with probability $p=\bar{k}/(N-1)$. The degree distribution is approximately Poisson, $p_k \simeq e^{-\bar{k}}\bar{k}^k/k!$, giving $\sigma_k\approx\sqrt{\bar{k}}$.
(iii)~\emph{Gamma network}: integer degrees drawn from a gamma-shaped distribution with mean $\bar{k}$ and standard deviation $\sigma_k$.
(iv)~\emph{Hamsterster}: the empirical friendship network described in Refs.~\cite{Kunegis2013,Feng2019}, with $N=2426$, $\bar{k}\approx 13.7$, and $\sigma_k\approx 19.9$.
Networks (i) and (iii) are created via the configuration model~\cite{Bollobas1980, StegerWormald1999} with repeated stub-matching until a simple graph is obtained (self-loops and multi-edges removed).

For the WT distributions, in addition to the gamma family described by Eq.~\eqref{gammadist}, we employ a log-normal parametrization with the same mean $r^{-1}$ and standard deviation $r^{-1}/\sqrt{\alpha}$.
The waiting time is drawn as $t = e^X$ where $X$ is sampled from a Gaussian with mean $-\ln r - s^2/2$ and standard deviation $s$, with $s^2 = \ln(1+1/\alpha)$.

\begin{figure}[h]
\includegraphics[width=\columnwidth]{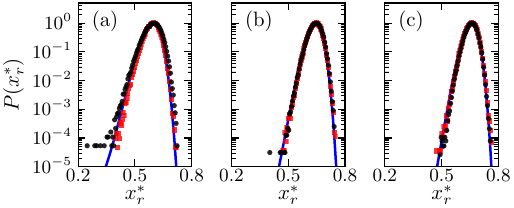}
\vspace{-7mm}\caption{\label{fig:recovery}Final outbreak-size distribution on a regular network with $N=3000$, $\bar{k}=10$ and  $R_0 = 1.5$, with gamma-distributed recovery and exponential infection. Here, the recovery shape parameter was (a)~$\alpha_{\rm rec} = 0.8$, (b)~$\alpha_{\rm rec}=1.5$, and (c)~$\alpha_{\rm rec}=2.0$, each normalized by its extensive-outbreak peak.
Black circles: non-Markovian network simulations.
Blue line: WKB theory at $R_0^{\rm eff}$ from Eq.~\eqref{eq:R0_eff}.
Red squares: well-mixed Markovian simulation for the same $R_0^{\rm eff}$.}
\end{figure}

\appsection{Non-exponential recovery}
\label{app:recovery}
In this Appendix, we  show that the transmissibility-based mapping applies equally to non-exponential recovery.
For gamma-distributed recovery with shape $\alpha_{\rm rec}$ and exponential infection, Fig.~\ref{fig:recovery} shows that the WKB theory with $R_0^{\rm eff}$ from Eq.~\eqref{eq:R0_eff} captures the full outbreak-size distribution across a wide range of $\alpha_{\rm rec}$.
Decreasing $\alpha_{\rm rec}$ below unity (more dispersed recovery) decreases $T$ and suppresses the outbreak, opposite to the effect of broader-than-exponential infection, confirming the opposite role of $f_i$ and $\Phi_r$ in Eq.~\eqref{eq:T}.

\appsection{Degree heterogeneity at fixed mean outbreak}
\label{app:heterogeneity}

\begin{figure}[h]
\includegraphics[width=\columnwidth]{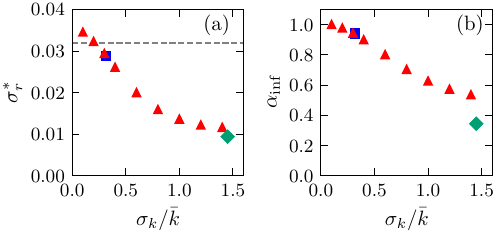}
\vspace{-7mm}\caption{\label{fig:std_vs_cov}Effect of degree heterogeneity at fixed conditional mean $\overline{x_r^*}\approx 0.6$, for various networks with $N=3000$ and $R_0=1.5$, with gamma WTs for infection and exponential recovery.
(a)~Conditional standard deviation $\sigma_r^*$ versus degree coefficient of variation $\sigma_k/\bar{k}$.
(b)~Infection-shape parameter $\alpha_{\rm inf}$ required to achieve $\overline{x_r^*}\approx 0.6$ versus $\sigma_k/\bar{k}$.
Blue squares: ER with $\bar{k}=10$; red triangles: gamma network with $\bar{k}=10$ and varying $\sigma_k$; green diamond: Hamsterster.
Dashed line in (a): WKB prediction from Eq.~\eqref{eq:sigma}.}
\end{figure}

To isolate the effect of degree heterogeneity on outbreak fluctuations, here we fix the conditional mean outbreak fraction $\overline{x_r^*}$ and examine $\sigma_r^*$ as a function of $\sigma_k/\bar{k}$---the coefficient of variation of the  degree distribution, see Fig.~\ref{fig:std_vs_cov}.
As network heterogeneity increases, maintaining the same $\overline{x_r^*}$ at fixed $R_0$ requires smaller $\alpha_{\rm inf}$, which increases the probability of rapid infections and narrows the outbreak-size distribution.
The resulting $\sigma_r^*$ decreases relative to the well-mixed WKB prediction, confirming the systematic deviation from the universal collapse described in the main text.
The Hamsterster network deviates from the trend of the gamma-distributed one in both $\alpha_{\rm inf}$ and $\sigma_r^*$, consistent with the role of degree-degree correlations, which effectively increase heterogeneity~\cite{Newman2002assort,Korngut2025Assaf}.

\appsection{Bond percolation on degree-correlated networks}
\label{app:assortative}

Here we extend the bond-percolation theory to networks with degree-degree correlations.
The uncorrelated percolation theory assumes that the degree of a neighbor reached by following an edge is independent of the degree of the source node.
Real networks, however, exhibit degree-degree correlations: the conditional probability $P(k'|k)$ of finding a degree-$k'$ node at the end of an edge from a degree-$k$ node differs from the uncorrelated form $k'p(k')/\bar{k}$~\cite{Newman2002assort}.

For each degree $k$, let $\theta_k$ be the probability that a given neighbor of a degree-$k$ node does \emph{not} transmit the disease to it.
A neighbor of degree $k'$ fails to transmit if either the bond is unoccupied (probability $1-T$) or the bond is occupied but that neighbor was never infected through its remaining $k'-1$ contacts, giving the self-consistency equation~\cite{VazquezMoreno2003}
\begin{equation}\label{eq:theta_k}
\theta_k = \sum_{k'} P(k'|k)\,\bigl[1-T+T\,\theta_{k'}^{\,k'-1}\bigr].
\end{equation}
A node of degree $k$ escapes infection if and only if none of its $k$ neighbors transmit to it, so the mean outbreak fraction is
\begin{equation}\label{eq:xr_corr}
\overline{x_r^*} = 1 - \sum_k p_k\,\theta_k^{\,k}.
\end{equation}
In the uncorrelated limit $P(k'|k) = k'p(k')/\bar{k}$, all $\theta_k$ collapse to a single $\theta$ and Eqs.~\eqref{eq:perc_gen} are recovered.
For the correlated Hamsterster network, we solve Eq.~\eqref{eq:theta_k} numerically and obtain $\overline{x_r^*}$ from Eq.~\eqref{eq:xr_corr}; the effective Markovian $R_0^{\rm net}$ from Eq.~\eqref{eq:R0_net} is then defined as the value reproducing this $\overline{x_r^*}$ in a Markovian simulation on the same network.

\begin{figure}[h]
\includegraphics[width=\columnwidth]{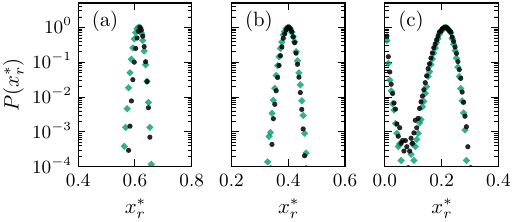}
\vspace{-7mm}\caption{\label{fig:hamsterster_dist}Outbreak-size distribution on the Hamsterster network with $N=2426$, $\bar{k} \approx 13.7$ and $R_0 = 3$, for gamma WTs for infection and exponential recovery. In (a-c) the infection shape parameter equals: $\alpha_{\rm inf} = 0.4$ (a), $\alpha_{\rm inf} = 0.8$ (b) and $\alpha_{\rm inf} = 1.2$ (c).
Black circles: non-Markovian network simulations.
Green diamonds: Markovian ($\alpha_{\rm inf} = 1$) simulations on the same network for $R_0^{\rm net}$ from Eq.~\eqref{eq:R0_net}.}
\end{figure}

Figure~\ref{fig:hamsterster_dist} shows the full outbreak-size distribution on the Hamsterster network for various gamma-distributed infection WTs and exponential recovery.
As in Fig.~\ref{fig:heterogeneous}, which reports mean and standard deviation, the non-Markovian simulations agree well with effective Markovian simulations on the same network at $R_0^{\rm net}$ from Eq.~\eqref{eq:R0_net}.
This confirms that the equivalence mapping captures the entire distribution shape on this strongly correlated, empirical topology.

\end{document}